\documentstyle[prl,aps,floats]{revtex}

\begin{document}
\def\b{\begin{equation}}
\def\e{\end{equation}}
\def\s{\vskip 0.4cm}
\draft
\vskip 0.5cm
\title
{On the Final State of Spherical Gravitational Collapse}

\author{Abhas Mitra}

\address{Nuclear Research Laboratory, Bhabha Atomic Research Center,\\
Mumbai-400085, India\\ E-mail: amitra@apsara.barc.ernet.in}


\maketitle

\begin{abstract}
Following our  recent finding\cite{1} that for the final state of
continued spherical gravitational collapse of sufficiently massive bodies,
 the final gravitational mass of the
fluid, $M_f\rightarrow 0$, we show that  for a physical fluid the eventual
value of $2M_f/R_f\rightarrow 1$ rather than $2M_f/R_f <1$ (the speed of
light $c=1$ and  the gravitational constant $G=1$), indicating the approach
to a zero-mass black hole. We also point out that as the final state is
approached the curvature components tend to blow up, also the proper
radial distance $l$ and the proper time (measured along a radial
worldline) $\tau \rightarrow
\infty$  indicating that actually the singularity is never attained.
We also identify that, the final state may correspond to the local 3-speed
attaining either $V=0$ or
$V\rightarrow c$, even though invariant circumference contraction speed $U=
dR/d\tau \rightarrow 0$. Nonetheless, at a finite
observation epoch,
 such Eternally Collapsing Objects (ECOs)
may have a local speed of collapse $V \ll c$ and the labframe speed of collapse
may be negligible because of high surface gravitational redshift. However,
if quantum back reaction in the strong gravity regime would cause
 a phase transition
of the form,
 pressure $p= -\rho$, where $\rho$ is the density of the collapsing fluid, it
  may be possible to have static Ultra Compact Objects (UCOs)
of arbitrary high mass\cite{2}. While
supposed Black Holes have no intrinsic magnetic field, ECOs or UCOs are likely to
possess strong intrinsic magnetic field and we point out that there are
already some tentative evidence for existence of such intrinsic magnetic
fields in some Black Hole Candidates\cite{3,4}.  For the benefit of the readers who may not have
gone through Paper I, we also include  here the summary of  the same. It clearly shows
that the central result of Paper I  can be derived even without knowing the meaning
of the nomenclature $V$ or without imposing any of  property of $V$ such as whether
$V <1$ or not. In addition we  consolidate the same result from other physical considerations too.\footnote{Accepted for Publication in Foundation of Physics Letters}

\end{abstract}


\section{Introduction}
One of the  questions haunting the physics since Newton unravelled the
universal attractive nature of Gravity is what would be the ``final state''
of spherical gravitational collapse of sufficiently massive bodies. As of
now, it is generally believed that collapsing bodies whose eventual mass
is more than $\sim 4-5 M_\odot$ (solar mass), must become black holes (BH).
Here, it must be mentioned that, in a strict sense, it has never been
possible to find any {\em exact} analytical solution of the collapse
problem by invoking an exact equation of state (EOS) and radiation
transport properties of the collapsing fluid. A deep introspection would
show that, neither will it ever be possible to do so in a strictly rigourous manner.
This is so because, in order to handle the collapse problem in a really
rigourous manner, without invoking any preconceived notions, one would require
to know the march of the {\em ever evolving} EOS and radiation transport
properties all the way upto the singularity! One can claim that one needs
to know the march of the EOS only upto the state of formation of the
horizon because, it can be shown that once the collapsing matter is
enshrouded by an Event Horizon (EH), collapse upto the central singularity is inevitable.
But since the mass of the collapsing fluid ($M$) is changing, so is the
would be location of the EH ($R_g=2M$), and hence it is not known beforehand.
 Thus,
in principle, one  would indeed require to know the march of relevant
parameters all the way upto the singularity in order to study the collapse problem
in an exact fashion.
At this juncture, one may recall that as far as analytical solutions are concerned,
the faith in the inevitability of formation BHs or
EHs essentially relies on the analytical solution of Oppenheimer and
Snyder (1939, OS)\cite{5}. OS considered a highly idealized case of collapse of a
pressureless ``dust'' of uniform density. But we have already shown that,
if interpreted correctly and consistently, their work shows that no
trapped surface or Event Horizon (EH) is formed at a finite value of
$M$\cite{1}.
Further, starting from the seventies, several authors have tried to
semi-analytically examine  the case of collapse of a relatively more
realistic fluid - an ``inhomogeneous dust''. These authors claim that
the collapse of an inhomogeneous dust need not lead to the formation of an
EH ! In other words, inhomogeneous collapse may lead to the formation of
the so-called ``naked singularities''\cite{6}. Again, here, one may invoke the
so-called ``cosmic censorship conjecture'' (CCC) of Penrose\cite{7}
which asserts that
for realistic cases, collapse must lead to the formation of BHs rather
than naked sigularities. However, given an initial gravitational mass
$M_i$, the running value of the gravitational mass, $M$, would be constantly
changing until the EH is formed and CCC {\em can not tell at what value of
$M$, the horizon would form}. Such questions can be hoped to be answered
only by the solution of actual collapse equations, if it were possible.
Thus, even if one accepts CCC for the time being, in order to rigourously
study the problem of gravitational collapse, one indeed needs to know the
evolving EOS for arbitrary high density and temperature of the collapsing
fluid. Second, since there is no proof for CCC, and it being what it is, a
conjecture, we realize that  it is unrealistic to hope for any strictly
rigourous solution of
the  problem of collapse by any numerical means.

And it is in this perspective that recently we have discovered an
extremely important {\em global property} of the General Relativistic (GR)
collapse equations for a spherically symmertic physical fluid which means
that (i) no trapped surface is ever formed and if the  fluid indeed
undergoes continued collapse, the final state must have $M_f \equiv 0$ as
$R_f=0$\cite{1}. In view of the importance of this result and for the sake
of those readers who may not have read Paper I, we present the salient
feature of it below. After this we shall proceeed to address the aspects
listed in the abstract of this paper.

\section{Crux of Paper I}
By analyzing the GR collapse equations in a generic  comoving
coordinate system  described by the interior
metric:
\begin{equation}
ds^2= g_{00} dt^2 + g_{rr} dr^2 + R(r,t)^2
(d\theta^2+\sin^2\theta d\phi^2)
\end{equation}
where $\theta$ and $\phi$ are angular coordinates,  $R$ is the circumference
coordinate and $N(r)$ is the total number of baryons
 inclosed within the sphere of constant coordinate radius $r$,
we have shown in  paper I\cite{1},  that
\b
{2 M(r,t)\over R} \le 1
\e
This result depends solely on the global properties of the field equations
manifest through the inverse curvature like parameter:
\begin{equation}
\Gamma \equiv {dR\over dl}
\end{equation}
where  $dl= \sqrt{-g_{rr}} dr$ is an element of
proper radial distance along the worldline and
the contraction (or expansion) rate of the invariant circumference radius
 is\cite{8,9,10,11,12,13,14}
\begin{equation}
U\equiv dR/d\tau,
\end{equation}
Here $d\tau= \sqrt{g_{00}} dt$ is an element of proper
time measured by a local $r=constant$ observer. Clearly these two parameters
are interrelated through an equation
\begin{equation}
U=\Gamma V
\end{equation}
where
\begin{equation}
V \equiv {dl\over d\tau} = { \sqrt{-g_{rr}} dr\over \sqrt{g_{00}} dt}
\end{equation}
Though the actual collapse equations are extremely complicated and
nonlinear and defy
{\em exact}  solutions either analytically or numerically,
nevertheless, they are interwined globally through an Eq.
connecting\cite{8,9,10,11,12,13,14}
 $\Gamma$ and $U$:
\begin{equation}
\Gamma^2=1+U^2 -{2 M\over R}
\end{equation}
Eqs.(5) and (7) can be combined in a master global equation:
\begin{equation}
\Gamma^2 (1-V^2) =1 -{2  M\over R}
\end{equation}
Or,
\begin{equation}
{1\over - g_{rr}} (\partial R/\partial r)^2 (1-V^2) =1 -{2  M\over R}
\end{equation}
Note that although $\Gamma$ has a partial derivative nature with respect
to $r$, it is a total derivative with respect to $l$\cite{10,11} because the notion of
a fixed $t$ is implied in the definition of $dl$.
Also, note that for radial worldlines with $d\theta =d\phi=0$,
 Eq.(1) can be rewritten as\cite{15}
\begin{equation}
ds^2= dt^2 g_{00} [ 1-V^2]
\end{equation}
Since for material particles, $ds^2 \ge 0$, the foregoing Eq. demands that
\begin{equation}
g_{00} (1-V^2) \ge 0
\end{equation}
Further since the determinant of the metric tensor $g= R^4 \sin^2 \theta
g_{rr} g_{00}$ is always negative\cite{15}, $g \le 0$, it follows that
\begin{equation}
g_{rr} g_{00} \le 0, ~or,~ -g_{rr} g_{00} \ge 0
\end{equation}
By dividing the inequality (11) by (12), we obtain
\begin{equation}
 {1-V^2\over -g_{rr}} \ge 0
\end{equation}
By feeding the inequality (13) in Eq.(9) we find that both sides of it are positive.
Therefore, the collapse equations obey the most general constraint that
\begin{equation}
1- {2M\over R} \ge 0
\end{equation}
and from which Eq.(2) follows. This
 means that contrary to the intuitive idea\cite{16}, GTR actually does not allow
formation of trapped surfaces (there is a popular misconception that the ``singularity
theorems'' ``prove'' the existence of trapped surfaces, but actually, they
{\em assume} the existence of the same). This same result can also be
obtained by using the radiative Vaidya metric\cite{17}.
 Further note that in order that Eq.(2) is really satisfied, under the
assumption of positivity of mass\cite{18}, we must have
\begin{equation}
M \rightarrow 0~~as~~ R\rightarrow 0
\end{equation}
which means that the final gravitational mass of any gravitational singularity
must be zero, $M_f=0$.

 The fact $M_f=0$ means that the ``gravitational mass defect'' $\rightarrow
M_i c^2$, the original mass-energy of the fluid. As is well known, as any
self-gravitating isolated body contracts it becomes hotter and radiates
energy (in the form of photons, neutrinos etc.)
and this is known as Kelvin-Helhmlotz process. This result is
obtained by just combining classical Virial
Theorem and total energy conservation. Therodynamically,
this happens because of negative specific heat associated with self-gravity.
As the self-gravitating body contracts it becomes gravitationally more
bound, i.e, its total energy decreases. Therefore it must  radiate out in
order to
satisfy total energy conservation. In relativity loss of energy from the body
means decrease of its gravitational mass $M$.
 Note that this loss/decrease of
mass is not due to loss of physical matter from the body and the baryonic
mass of the body, $M_0$, can remain fixed while $M$ decreases steadily.

 On the other hand if we focus our atention on the fixed baryonic mass of
the system,
 it may indeed be possible to
pack baryons sufficiently closely to achieve a state $2M_0/R >1$ and in
fact to chase the limit $2 M_0/R\rightarrow \infty$. In Newtonian physics
the gravitational mass $M\equiv M_0$, and thus, Newtonian physics may
admit of a BH. It is small wonder then that {\em the concept of a BH
actually arose almost two hundred years ago}\cite{19}. And the idea that
the formation of a  ``trapped surface'' with $2GM/R >1$ is most natural is
thus deeply ingrained into the intuitive (Newtonian) notions which do not
distinguish between baryonic  and gravitational mass. In
contrast, in GTR,  the total mass energy $M \sim M_0 +E_g + E_{\rm
internal}+ E_{\rm dynamic}$; where the gravitational energy is negative,
and is some evolving nonlinear function of $M$, $E_g= -f(M, R)$. In the
limit of weak gravity $f \sim GM^2/R$, but as the collapse proceeds, the
grip of self-gravity becomes tighter and $M$ starts becoming reasonably
smaller   than $M_0$ or $M_i$. And if it were possible for $M$ to assume
negative values, for continued gravitational collapse, the non-linear
$E_g$ would relentlessly push $M \rightarrow -\infty$. However, there are
positive energy theorems\cite{18}, which state that the mass-energy of an {\em
isolated body} can not be negative, $M\ge 0$, and, physically, which means
that, gravitation can never be {\em repulsive}. And hence, the continued
collapse process comes to a {\em decisive end} with $M_f=0$.
Physically the state $M=0$ may occur
when the energy liberated in the process (as measured by a distant
inertial observer $S_\infty$), $Q\rightarrow M_i c^2$ because $M_f=M_i - Q$.

Further the
present study would show that the final $M_f=0$ state corresponds to the
formation of a horizon at $R_f=0$ with $2GM_f/R_f=1$. And since horizon is
formed at $M_f=0$ the system cannot lose further energy and its mass
cannot become negative. Thus the result that horizon would be formed at
$M_f=0$ substantiates the positive energy theorems.
Our result that the assumtion of formation of ``trapped surfaces''
 is not
realized confirms Einstein's idea\cite{20} that Schwarzschild singularities can not
occur in practice. However, Einstein's exercise was based on the non-collapsing
scenarios and the theoretical formalism for GTR collapse including
pressure was developed in the sixties, much after Einstein's exercise.

One important thing to note here is
that in the above derivation of Eq.(2) or (14),
 {\em we did not even require to know
the meaning} of the nomenclature $V$,  and neither  did we
 require to impose any condition on the property of $V$, (e.g, whether
 $V \le 1$ or not).

 \subsection{Physical Interpretation of $V$}
 Landau \& Lifshitz (LL)\cite{15} first defined  local 3-speed as
$dl/d\tau$ for a
``particle'' moving in a ``{\em constant}'' gravitational field, i,e, when
metric components do not depend on time ``$x^0$'' or ``$t$'' e.g., a
particle moving around a spherical body. Some readers might think that,
the concept of 3-speed, therefore, may not be extended to a dynamic case,
i.e, when the metric components are functions of $x^0$. This is not true
because (i) both $dl$ and $d\tau$ remain definable even when $g_{ik}$s depend
on $x^0$ \cite{15} and (ii) then one would not even be able to define the
energy momentun tensor of a fluid. For instance, in the Section entitled, ``The centrally
symmetric gravitational field'' LL\cite{15} discuss the field equations
of a dynamic fluid involving local 3-speed (below eq. 100.20).
In our case,  $V$ happens to be the
 local 3 speed of an  fluid element at a paticular $r$ because of {\em
implicit} dependence of $r$ on $t$.
 At first sight it may appear
that since $r$ is a
comoving coordinate, $dr \equiv 0$, and therefore $V\equiv 0$.
 Then how to extract the {\em implicit}
dependence of $r$ on $t$? As explained in Paper I, it should be done in
the following way:

  Focus attention on a certain comoving observer at $r=r$ at its comoving time $t=t$.
Let this be  the observer 1. There is an adjacent observer 2 at $r =
r+dr$ at $t=t$ as recorded by clock 1.
Take a snapshot of this scenario at a time $t =t$ as recorded by clock 1.
Now take another snapshot of the fluid layers at a time $t=t+dt$ as
recorded by the same clock 1. Now superimpose the two snapshots. Observer
1 would find that his present position (at $t= t+dt$) is matching with the
previous position held by clock 2 at $r=r+dr$. From this measurement,
observer 1 would know that during the interval $dt$ as recorded by
himself, he has travelled a distance $dr$ {\em with respect to the
previous position of observer 2}.
 This is the significance of $dr$ in the comoving metric (1).
And since in this sense, $dr \neq 0$, $V \neq 0$ for a dynamic fluid:
\b
V= {\sqrt{-g_{rr}} dr\over \sqrt{g_{tt}} dt}
\e

Note that in the foregoing Eq. $dr$ is the distance between two
neighbouring layers as recorded in the first snapshot, i.e, $dr$ is
measured at a fixed $t$. And in Eq.(16), $dt$ is measured solely by the 1st clock, i.e, by
 a clock fixed at $r=r$. Therefore, the $dr/dt$ term appearing in Eq.(16)
can actually be written as
\b
V=  {\sqrt{-g_{rr}} dr\mid_t \over \sqrt{g_{tt}} dt\mid_r}
\e
 Also note that since
\b
 {\partial R\over \partial r} = { d R\over d r\mid_t}
\e
and
\b
{\partial R\over \partial t} = {dR\over dt\mid_r},
\e
 the $dr/dt$ appearing in Eq.(17) may be expressed as
\b
{dr\over dt} = { (\partial R/ \partial t)\over (\partial R/ \partial r)}
\e

Now by using Eqs.(3), (4) and (20),
we find that Eq.(16) can be further rewritten as $V= U/\Gamma$.
 In this way we have a
complete explanation as to how Eq.(5) is valid even though $\Gamma$ and $U$
involve partial derivatives.

 However speed of a given fluid element measured by an observer 1
 comoving with the same
fluid element is of course zero {\em unless} the measurement is done with
respect to the adjacent observer 2, $V_{com}\equiv 0$. This is true in Newtonian
hydrodynamics too. So one must distinguish between these two notions of
comoving speed. When this latter view is adopted the comoving metric will be
\b
ds^2 = g_{tt} dt^2
\e
rather than Eq. (1); and corresponding $V_{com} \equiv 0$. We may recall that
in the concept of the 4 velocity of the fluid in the comoving frame $u^\mu$, one tracks
a given fluid element irrespective of its relative displacement from a
neighbouring element and hence the spatial part of $u^\mu$ involves $V_{com}=0$
rather than $V$.
 While we explain the physical meaning of $V$ we remind the reader again that
 in the derivation of Eqs. (2) and (14) {\em we did not even require to know
 what the nomenclature $V$ stood for}, neither  did we  require to impose the
condition that $V \le 1$. For previous use of the concept of speed in the
context of comoving coordinates see Eqs. 2.79-80 in ref.\cite{21} and Eqs.
6.42-45 in ref.\cite{22}.

\section{Completely Independent Proofs}
It is likely that some readers would just refuse to accept the derivation which led to Eq.
(2) even if he/she would not be able to point out any specific error in it.
Let us remind them that,
 our basic result that there cannot be any horizon at a finite
value of $R$ can be obtained from several other independent considerations
too:
\subsection{Radial Geodesic Around a Schwarzschild BH}
 Let us assume, for the time being, the existence of a Schwarzschild BH
with finite mass $M$ and horizon size $R_g = 2M$.  The spacetime for $R \ge R_g$ is
described by
\b
ds^2 = dT^2 ( 1- 2M/R) - {dR^2\over (1- 2M/R)} - R^2 (d\theta^2 + \sin^2
\theta d\phi^2)
\e
For radial geodesics of test particles, $d\theta= d\phi =0$, and it
follows that along the radial geodesic\cite{23}
\b
\left({dR\over dT}\right)^2 =   {(1-2M/R)^2\over E^2} [E^2 - (1-2M/R)]
\e
where $E$ is the conserved energy per unit rest mass.
As $R\rightarrow 2M$, the foregoing Eq. shows that
\b
\left({dR\over dT}\right)^2 \rightarrow (1-2M/R)^2
\e
By transposing  this Eq., we find that
\b
dT^2 (1-2M/R) \rightarrow {dR^2\over 1- 2M/R} = dz^2~~ (say)
\e
By inserting the above relation in Eq.(22), we find that, for a radial
geodesic,  as $R\rightarrow 2M$
\b
ds^2 \rightarrow dz^2 - dz^2 \rightarrow 0
\e
Note that this is an {\em exact and straightforward result} and nothing can be
done to undo it.

  This means that, the {\em timelike geodesic} would turn  null {\em if an
EH would exist}. But this is not possible if EH is really a regular region
of spacetime and it demands that there cannot be any finite mass Sch. BH
with a  finite horizon ($R_g=2M >0$). But if we still insist on having a BH, the only honourable
solution here would be to recognize that the EH is actually a true
singularity, i.e, the central singularity. This would mean that $R_g =0$
or $M=0$. So if we assume that there is a BH, the only admissible value of its
mass is $M=0$ or else we may have finite mass objects which could be very
similar to BHs in many ways, but which are not exactly BHs.
Also note that from Eq.(24),  eventually, at $R=2M$ we have
\b
{dR\over dT} \rightarrow 0;~~ dR \rightarrow 0
\e
because by definition $dT$ is an infinetisimal. At this point, some of the
readers may have one confusion: It is well known that  $\Delta T_g=
\int_R^{R_g} dT =\infty$, and some readers may confuse $\Delta T_g$ with
$dT$ and may errorenously think that $dT =\infty$. The fact that $dR=0$ at
$R=2M$ indicates that the EH is the true end of the spacetime or it is the
central singularity $R=0$.

\subsection {Same Result Using Kruskal Coordinates}
If $u$ and $v$ are the Kruskal coordinates
the corresponding metric is\cite{1,23}
\begin{equation}
ds^2 = {32 M^3\over  R} e^{-R/2M} (dv^2 -d u^2)
\end{equation}
and
over the entire  Kruskal diagram, we have\cite{23}
\b
u^2-v^2= \left({R\over 2M} -1\right) e^{R/2M}
\e
so that
\b
 u^2 = v^2; \qquad v/u =\pm 1, \qquad R= 2M
\e
By differentiating Eq.(29) by $T$, we find
\b
2 u {du\over dT} - 2v {dv\over dT} = {e^{R/2M}\over 2M} {R\over 2M}
{dR\over dT}
\e
Using  Eq.(27) in the
foregoing Eq., we find that the R.H.S. of Eq.(31) is zero at $R=2M$ and
\b
{du\over dv} \rightarrow {v\over u}; ~~R \rightarrow 2M
\e
Now combining Eqs.(30) and (32), we  obtain
\b
(du/dv)\rightarrow  \pm 1;\qquad du^2 \rightarrow  dv^2,~~R \rightarrow 2M
\e
Hence the Kruskal metric
\begin{equation}
ds^2 = {32 M^3\over  R} e^{-R/2M} (dv^2 -d u^2) \rightarrow 0;~~ R
\rightarrow 2M
\end{equation}
Again note that this is an absolutely straight forward and exact result and nothing can
be done to undo it.  Yet some of the readers may develop a confusion here and
on the basis of which they might feel
that  the Eq.(34) could be incorrect because of the
following reason:
 If we divide both side of Eq.(28) by $ds^2$ we would obtain a first
integral of motion along a radial geodesic:
\begin{equation}
 {32 M^3\over  R} e^{-R/2M} [(dv/ds)^2 - (d u/ds)^2] =1
\end{equation}
anywhere including at $R=2M$. Eq.(35) may convey an impression that $dv^2
> du^2$ everywhere. But it would be incorrect to assume so in general. All
that this identity shows is that
\b
{dv^2 -du^2\over ds^2} >0
\e
If there is a fraction $z= y/x >0$, it does not necessarily mean $y >0$,
$x >0$ because
 a $0/0$ form can also  be $>0$. Seen more clearly,
Eq.(35) is nothing but the following {\em identity}:
\b
{ds^2\over ds^2} =1
\e
And an identity like $x/x=1$ cannot determine the value of $x$ by itself,
all it means is that $x=x$ where $x$ can be both $>0$, $<0$ or $=0$.

There might yet another illogical attempt to deny the correctness of Eqs.
(26) and (34) in the following manner. Some readers may insist that Eq.
(23) is not extendible upto $R=2M$ at all because the $R, T$ coordinates
``break down'' there. But if the ``$R, T$'' coordinate system would really
break down, the Kruskal coordinates should also ``break down'' because
{\em Kruskal coordinates are not directly linked to any invariant, any
scalar or any observables, but on the other hand, they are built using
solely $R$
and $T$}. On the other hand, $R$ is directly related to {\em
invariant} circumference radius$/2 \pi$. Further, since the curvarure {\em
scalar} is $K= 48 M^2/R^6$, $R$ is also inversely proportional to the 1/6
th power of $K$. Similarly, $T$ appearing here is the {\em proper time}
measured by a distant observer $S_\infty$ and thus $T$ too has a solid physical
significance unlike $u$ and $v$. So if the EH or the region interior to it
are to be described by $u=u(R, T)$ and $v=v(R, T)$, by definition, $R$ and
$T$ must remain valid coordinates there. Thus Eq. (23) must remain valid
wherever $u$ and $v$ are definable even though Eq. (22) may not be valid.

  Had we differentiated Eq.(29) by $v$ or $R$ instead of $T$ we would have
obtained the same result because it can be seen that
\b
{dv\over dR} = {rv\over 8M^2} (R/2M -1)^{-1} + {u\over 4M} {dT\over
dR}= \infty;\qquad R\rightarrow 2M
\e
GTR demands that the geodesic for a material particle must be timelike,
$ds^2 >0$ at any regular non-singular region of spacetime. But we found
that, on the horizon, we would have $ds^2=0$ along the geodesic. Since
this would be a violation of GTR it means that the assumption of existence
of a finite mass BH and its non-singular horizon ($R_g=2M >0$) is
incorrect.  Mathematically, if a BH would exist, its mass must be $M=0$ so
that the EH, $R_g=2M=0$, is a true physical singularity and not a mere
coordinate singularity.

 Also the occurrence of $ds^2=0$ {\em along a radial geodesic} must not be
confused with the fact that the EH is a null hypersurface.  In the 1D
Minkowski spacetime diagram, the 45$^\circ$ lines constitute the null
surface and it does not automatically mean that a timelike geodesic
approahing  these null lines would turn null.

So far, as noted by the present author, the only scientific critique to
our above result is due to Tereno\cite{24,25,26,27} who insisted that
$\mid du/dv\mid <1$ ar $R=2M$ (Tereno like us also justifiably used
Eq.(23) over the entire spacetime). But in Eq.  (33), we have already
found that $du/dv$ is indeed $\pm 1$ at $R=2M$. Then why did Tereno
apparently obtain a differnt result? This is because Tereno chose not the
simpler way of using Eq.(29)  to obtain $du/dv$. On the other hand, he
chose to find $du/dv$ directly\cite{1,24,25}: \b {du\over dv} = {u
-vE/\sqrt{E^2-1 +2M/R}\over v- uE/\sqrt{E^2 -1 +2M/R}}
\e
While on the EH, $u_H =\pm v_H$, the term $x=\sqrt{E^2- 1 +2M/R }= \pm E(1-
\epsilon/2E^2)$ close to the EH with $\epsilon = 1- 2M/R \rightarrow 0$.
Eq.(39) thus can assume 4 set of values:

Case I:  $x= + E(1-\epsilon/2E^2)$; $u_H = -v_H$, $du/dv = -1$

Case II: $x= + E(1-\epsilon/2E^2)$; $u_H = +v_H$,
\b
{du\over dv} = { (u_H/E^2)- (e/u_H) \over  (u_H/E^2) + (e/ u_H)}
\e
Case III:  $x= + E(1-\epsilon/2E^2)$; $u_H = +v_H$, $du/dv = +1$

Case IV: $x= + E(1-\epsilon/2E^2)$; $u_H = -v_H$ and

\b
{du\over dv} = {(e/u_H) - (u_H/E^2)\over (e/u_H) + (u_H/E^2)}
\e
While Cases I and III do show that $\mid du/dv\mid =1$ as clearly seen before,
Cases II and IV may lend an impression that, at the same time, $\mid
du/dv\mid$ could be $< 1$.
If this would happen, $ds^2$ would be both $0$ and
 $>0$ ar $R=2M$. But there must be a unique value and sign of $ds^2$,
and therefore, we must have
 $u_H^2=v_H^2$ equal to either $0$ or $ \infty$.

\subsubsection{Explicit form of $u_H$ and $v_H$}
To obtain the explicit form of $u$ and $v$ for a radial geodesic
in terms of only $R$, we need to
 recall the relationship between $T$
and $R$  which, in turn,
is obtained by integrating Eq.(23)\cite{23}:
\begin{equation}
{T\over 2M} = \ln{(R_i/2M-1)^{1/2} + \tan{(\eta/2)} \over (R_i
/2M-1)^{1/2} - \tan{(\eta/2)}} + \left({R_i\over 2M}-1\right)^{1/2}
\left[\eta + \left({R_i\over 4M}\right)(\eta +\sin \eta)\right]
\end{equation}
Here the test particle is assumed to be at rest at $R=R_i$
at $T=0$ (or for dust collapse, the starting point) and the
``cyclic coordinate'' $\eta$ is defined by
\begin{equation}
R= {R_i\over 2} (1 + \cos \eta)
\end{equation}
The initial value of $R=R_i = 2M/(1-E^2)$.
We find from Eq.(42) that, as $R\rightarrow 2M$, the logarithmic
term blows up and $T\rightarrow \infty$, which is a well known
result.
Correspondingly, as one enters the EH,
one would have
\begin{equation}
\cosh{T\over 4M} \rightarrow \sinh{T\over 4M} \rightarrow
{e^{T/4M}\over 2} =\infty
\end{equation}
Then it can be found that
\begin{equation}
u_H^2 =v_H^2 = 4e(1- 2M/R_i) \exp\left\{ \left({R_i\over 2M}-1\right)^{1/2}\left[\eta_H+
(R_i/2M)(\eta_H
 +\sin{\eta_H})\right]\right\}
\end{equation}
where
\begin{equation}
{\eta_H} =\arccos{(4M/R_i -1)}
\end{equation}
Since $R_i > 2M$,  ${u_H}^2 >0$. And in terms of $E$, we have
\b
u_H^2=v_H^2 = 4e E^2 \exp \left( \eta_H + {\eta_H + \sin \eta_H\over 2\sqrt{1-E^2}}\right)
\e
This shows that for $E=1$ ($R_i = \infty$), $u_H^2= v_H^2=\infty$ for any pre-assumed value of
$M$ as was concluded above. But are $u_H$ and $v_H$ finite for $E <1$ ? If
so, again, we would have both $ds^2=0$ and $\neq 0$ at $R=2M$. But since
this is not possible, we must have indeed have $u_H =v_H =\infty$, which,
by virtue of Eq.(45) means that, for consistency, we must have $M=0$.

\subsection{Physical Interpretation}
Both the Sch. metric (22) and Kruskal metric (28) can also yield an Eq.
like that of (10) and from which, we would see that,
  occurrence of $ds^2=0$ on the EH physically means occurrence of either
   $g_{00}=0$
or $V=1$ or both of them. And it is already well known that in
Sch. coordinate the local 3 speed of a free particle becomes equal to speed
of light at $R=2M$ because\cite{1,23}
\begin{equation}
V_{sc}^2 = 1- {(1- 2M/R)\over E^2} = 1;~~ R=2M
\end{equation}
 As mentioned by Landau \& Lifshitz\cite{15} any arbitrary metric
associated with the gravitational field of single body can be written
as
\b
ds^2 = g_{00} d{x^0}^2 (1-V^2)
\e
where $V$ is appropriately defined 3-speed.
Therefore, Eq.(48) itself could have been taken as a proof that $ds^2 (Sch) = 0$ at
EH (along a radial geodesic) because $g_{00}=g_{TT}  \neq \infty$ at
$R=2M$ (on the other hand $g_{TT} =0$ too at $R=2M$)
and of course $dx^0$ being an infinetisimal.

In GTR, like in Sp. Theory of
Relativity, the velocity addition law is such that once any observer
measures a particular speed $V=1$, {\em all other observers (coordinates) too
measure the same speed}  and this is the reason that all observers measure
speed of light $=c=1$ in GTR too.  Thus the speed of the test particle at
the EH
as measured by the Kruskal coordinates must also be $V_{Kr} =1$. The
3-speed in terms of Kruskal coordinates is
\begin{equation}
V_{Kr} \equiv {dl\over d\tau} = { \sqrt{-g_{uu}} du\over \sqrt{g_{vv}} dv}
={du\over dv}
\end{equation}
because in Eq.(28), $g_{vv} = -g_{uu}$.
By using Eq.(33), we   find that indeed $V_{Kr} =1$ at $R=2M$.  Further since
$ds^2$ is an invariant, we must have  $ds^2 (Kruskal)=0$ (at $R=2M$) once
$ds^2 (Sch) =0$ (at $R=2M$). Thus Eqs.(26) and (34) are fully self consistent.
Hence there
 is a complete
 physical as well as mathematical consistency between subsections A, B and
C.
Such an allround consistency would not have been possible had any of the
results contained in these  subsections been incorrect. In fact, now
{\em Tereno too has admitted} that, the local speed as defined by all the authors
becomes $c$ at $R=2M$ irrespective of the coordinate system used\cite{28};
(he has not cited his own previous preprints).
He has also correctly admitted that, therefore, one should have $V >c$
inside the EH\cite{28,29}. However, to avoid such a prospect, Tereno has
appealed that the basic concept of local speed as defined by Landau and
all other authors be changed\cite{28}!! It can be shown that as per his prescription,
all local speeds of free falling particles should be zero because they are
measured by another free falling observer situated at the same instantaneous location
(but the assumption of existence of a BH must be retained)!

  This section shows that for a {\em point mass} the    mathematical  solution (22) becomes physically
meaningful too only if $M=0$. This means that a BH of mass $M=0$ could be the
the endpoint of gravitational collapse. Then this is a hint that in our
constraint (2), physically, we should have $2M/R =1$ rather than $2M/R<1$.

\subsection{Same Constraint From OS Paper}
The Eq.(36)   of
Oppenheimer Snyder
(OS) paper\cite{5} looks like
\b
T \sim \ln{y^{1/2} -1\over y^{1/2} +1} + other ~ terms
\e
where $y= R/2M$ on the boundary of the collapsing star. In order that
the argument of the log term in the above Eq. is positive and $T$, the
{\em proper time} of a distant observer, is definable, it is necessary
that $y \ge 1$ or $2M/R \le 1$.
Thus the condition (2) that there is no trapped surface
 is very much steeped into Eq.(51).
Therefore final mass of the dust ball $M_f =0$ at $R=0$. But if the
``dust'' is (incorrectly) treated as a {\em coherent continuous physical fluid}
it would not be able to radiate anything including gravitational radiation
(because of assumed isotropy). In such a case the mass of the dust ball
must remain constant and hence even the initial mass of it $M_i=0$!
This
means that if we (incorrectly) insist that a spherical ``dust'' is treated as a
{\em coherent continuous physical fluid} the only allowed value of its
mass is $M=0$ and particle number $N=0$. Physically this means that if we strictly demand that a fluid
has no pressure at all $p\equiv 0$, it cannot really be considered as a
coherent continuous physical fluid.

 Thus a finite mass spherical ``dust'' with $p\equiv 0$ is merely
 a collection of {\em incoherent
 non-continuous particles}. In such a case there would be free space
between the ``dust'' particles and {\em even though the system is spherically symmetric,
the system is not isotropic}. Thus a spherical ``dust'' of $N$ particles is
an assembly of $N/2$ {\em incoherent} pairs of particles. In GTR any
mutually attracting pair of particles  emit gravitational radiation
unmindful of the presence of other incoherent pairs. Thus actually the
gravitational mass of a true incoherent non-continuous ``dust'' would vary during the process of collapse
like that of a physical fluid. And hence it is possible that as $R$ decreases,
$2M/R \le 1$ during its evolution. So whereas a spherical physical fluid (isotropic)
would  decrease its mass by emission of electromagnetic or neutrino radiation
a spherical (yet non-isotropic) ``dust'' would decrease its mass by
emission of gravitational radiation. But in either case $2M/R \le 1$.

Some readers may argue that (i) the work of OS, in any case, has been
redone in various form by various other authors and which show that
collapse of homogeneous dust results in a finite mass BH or, (ii) there should have been a modular sign in Eq.(36) of OS (which they
may have forgot to put!), and, when such a sign is introduced in Eq.(51), there would
be no problem in allowing a value of $y <1$ or in having $2M/R >1$.

 When one neglects the emission of (gravitational) radiation from
incoherent dust, its mass $M$ remains fixed, and if one starts with a
finite mass $M$ and ignores the inherent implication of Eq.(36) of OS (and
also ignore other inconsistenies discussed in Paper I), one can happily
conclude that trapped surfaces are formed and $2M/R >1$ as $R$ becomes
sufficiently small. The physical inconsistency in assuming the initial
condition of $M=finite$ would however always be present through the
modified $T-R$
relationship in any such study.

   In the absence of any pressure and neglect of
 radiation,  a given dust particle
undergoes free fall and behaves like a free test particle (note that
actually a pair of mutually accelerating particles do emit gravitational
radiation and this is ignored in this ``free particle'' scenario). In any
case, once one assumes this ``free particle'' view point, the physics of
dust collapse is somewhat similar to the radial inward motion of a ``free particle''
around a BH of mass $M$. The relation between $T$ and $R$ is  same as before and is
given by Eq.(42).

Since $\tan(\eta/2) = (R_i/R -1)^{1/2}$, we may rewrite Eq. (42)
in terms of a new variable
\begin{equation}
x = \left({R_i /2M -1\over R_i /R -1}\right)^{1/2}
\end{equation}
as
\begin{equation}
{T\over 2M} = \ln  {x +1\over
x-1} + \left({R_i \over 2M}
-1\right) \left[\eta + \left({R_i\over 4M}\right) (\eta + \sin \eta)\right]
\end{equation}
It can be easily found that, irrespective of $M$ being finite or zero, $T
\rightarrow \infty$ as $x \rightarrow 1$ or $R \rightarrow 2M$.
For the time being, we assume that $M \ge 0$. Then, from
Eq.(52), note that
\begin{equation}
x \le 1 ; \qquad for~ R\le 2M
\end{equation}
Thus for $R <2M$,
 $T$ is {\em not definable at all} because the argument of
logarithmic function in Eq.(53) can not be negative. This is similar to the problem
encountered in Eq.(51). However nobody seems to have pondered
 {\em whether the
situation which is leading to an imaginary $T$ is unphysical or
not}. As mentioned before, $T$ is the {\em proper time} measured by a
distant observer $S_\infty$ and thus $T$ cannot be imaginary.
  Note, as far as the distant inertial observer
 $S_\infty$ is concerned, he is either able
to watch an event ($T=finite$) or unable to do so ($T=\infty$).
Probably later authors realized that $T$ can not be
allowed to be imaginary, not because of the fact that $T$ is
still the {\em proper time} of a Galilean observer, a measurable
quantity, but because of the fact that, otherwise the esoteric new
coordinates, like $u$ and $v$ would not be definable. Thus we
find that a modulus sign was introduced in the $T-R$
relationship\cite{23}:
\begin{equation}
{T\over 2M} = \ln \mid {x +1\over
x-1}\mid + \left({R_i \over 2M}
-1\right) \left[\eta + \left({R_i\over 4M}\right) (\eta + \sin \eta)\right]
\end{equation}
But, the conceptual catastrophe actually becomes worse by this
tailoring. Of course, as the particle enters the EH, still, $T
\rightarrow \infty$. Note that, $T$ having become infinite,
definitely can not decrease, and more importantly {\bf inside the
EH, $T$ can not be finite} because, otherwise, {\em it would
appear that although the distant observer can not witness the
exact formation of the EH (in a finite time), nevertheless, he
can witness the motion of the test particle inside the EH}. This would
mean violation of causality and the existence of some sort of a
``time machine''. And we know that, it is only the comoving
observer who is supposed to witness both the formation of EH and
the collapse beyond it. But the foregoing equation tells that
 if $M >0$ (finite), the {\em the value of $T$ not
only starts decreasing} but also suddenly {\bf becomes finite} as
the boundary enters the EH ($R <2M$)!! And as the collapse is
complete, i,e.,
\begin{equation}
x =0; \qquad ~if~ M >0; \qquad at~ R=0
\end{equation}
 the corresponding value of $T=T_0$ required by the distant observer to
see the {\bf collapse to the central singularity within the EH} is simply
\begin{equation}
T = T_0 = 2M
 \left({R_i\over 2M}
-1\right) \left[\eta + \left({R_i\over 4M}\right) (\eta + \sin \eta)\right]
\end{equation}
By using Eq.(43), this can be rewritten as
\begin{equation}
T_0= \pi (R_i - 2M) \left( 1+ {R_i\over 4M}\right)
\end{equation}
And note that, if we really insist that $T_0=\infty$ as per the
original agenda, i.e, if we really insist that events occurring within the
EH are not visible to the distant observer, we must realize that $M=0$! This would mean (i)
there is no additional spacetime between the EH and the central
singularity, i.e, they are synonymous and the Sch. singularity is
a genuine singularity provided one realizes that by the time one
would have $R=2M$, the value of $M\rightarrow 0$, and (ii) the
proper comoving time for formation of this singularity is $\tau \propto
M^{-1/2} =\infty$. The latter means that, it is not formed at any
finite proper time, the collapse process continues indefinitely
and there is {\em no incompleteness for the timelike geodesics}.

\subsection{Final Confirmation: Acceleration Invariant}
There is yet another argument which corroborates the result obtained in
Paper I.  It is known that the radial component of acceleration ($a^R$, a
coordinate dependent quantity) acting on a test particle around a Sch. BH
blows up at $R=2M$.  Since $a^R$ itself is a physically measurable
quantity, its blowing up (eventhough it is coordinate dependent) should
have initiated introspection on the reality of occurrence of the horizon.
But conveniently this physical question is avoided with the hyperbole of
``coordinate singularity''. For the sake of argument, let us too accept it
for a moment.  But we can construct an acceleration INVARIANT/SCALAR (a
coordinate independent quantity)\cite{30,31,32,33}
  \b
a = \sqrt{a^i a_i} = {M\over R^2 \sqrt{1-2M/R}}
\e
 Note that {\em this scalar  blows up} at $R=2M$. Further, had there been a spacetime
beneath $R<2M$, it would have become imaginary there. Since $a$ is a
coordinate independent quantity now {\em we cannot get away by blaming it on
coordinate singularity}.

The only solution lies here is the admission of the fact that $M=0$, the
horizon is at $R_g=2M=0$ (the central singularity)
and there is no spacetime beneath the horizon.

\section{Schwarzschild or Hilbert Solution}
As they say, sometime truth is stranger than fiction:
It has been pointed out recently that the conventional Sch. solution is not
due to Schwarzschild at all, on the other hand, it is due to Hilbert.
The original paper of Schwarzschild\cite{34} where he worked out the spacetime around a
{\em point mass}, $M$ has recently been translated into English by Loinger
and Antoci\cite{31}. Even before this Abrams\cite{30}
 pointed out that, the original Schwarzschild
solution is
\b
ds^2 = dT^2 \left( 1- {2 M\over R}\right) - {dR_*^2\over \left( 1-
{2M\over R}\right)} -R^2 (d\theta^2 + \sin^2 \theta d\phi^2)
\e
where
\b
R= [R_*^3 + (2M)^3]^{1/3}
\e
The important point is that here $R$ is not the radial variable, on the
other hand the radial variable is $R_*$;
 the {\em point mass is sitting at} $R_* =0$ {\em and not at} $R=0$. Thus at,
 $R_*=2M$, obviously there is no singularity, whereas there is a central
singularity at $R_*=0$. Since there is no spacetime beneath $ R_* <0$,
there cannot be any spacetime beneath $ R < 2M$. The prevalent Sch. solution,
which is actually the Hilbert solution is, however, still a mathematically
valid solution for a ``point mass''. So the question asto which is the
physically valid solution, has to be decided on physical reasoning. Here,
some authors\cite{30,31,35,36,37} have rejected the Hilbert solution  because, according
to it, there would be a spacetime beneath $R <2M$ and the scalar $a$ would
then first blow up and then become imaginary. On this ground, these
authors, claim that the original Schwarzschild solution is the physically
valid solution. If so, in this picture, obviously, there would be no BH.
 On the other hand, there is a problem with the original Sch. solution; as
admitted by Sch. himself\cite{34}, in the limit of weak gravity, his equation does
not exactly reduce to a Newtonian form. On the other hand, we know that,
the prevalent Sch. solution, i.e, the Hilbert solution does yield the
correct Newtonian form. Further, the angular part of either Schwarzschild or
Hilbert solution (or  any spherically symmetric metric) shows that $R$
rather than $R_*$
is the Invariant Circumference Coordinate. Therefore, Hilbert solution cannot
 really be rejected. Then how do we resolve this paradox? We note that,
the problem lies with our premises that ``there is a {\em point mass} with
a finite gravitational mass''! And the solution lies in realizing that
 if there is a body of finite mass, it cannot be considered as a {\em
geometrical point} even at a classical level. On the other hand, if we insist that there is a point
mass, its gravitational mass must be zero. When $M=0$, Sch. solution and
Hilbert soln. become identical, $R_*\equiv R$, and there is no EH, but
there is only a
central singularity in a mathematical sense. One may note here that in
classical electrodynamics there are ``point charges''. And existence of
 ``point charges'' implies singularities. On the other hand, in quantum
field theories, there are no point charges, for instance, an electron is
actually not a ``point charge''. We find here that in GTR, even at a classical, non-quantum
level, there cannot be any ``point mass'' (having finite $M$).

\section{Class of Solutions}
Having confirmed the basic result obtained in Paper I by several
independent modes,
now we shall focus attention on the salient features of the final states
of spherical  gravitational collapse as dictated by Eqs.(5) and (8).
Before we do this,
see that
all the initial states may be considered to have $U_i=V_i=0$ and $\Gamma_i^2 =
1- 2 M_i/R_i$.

 We have found that, as to the
final state, the master Eqs.(5) and (8) admit basically four types of solution:

\subsection{Dust Solution}
 We discussed in Paper I that if a dust is (incorrectly) considered as a
coherent continuous physical fluid,
for the dust solutions, $\Gamma_f$ remains pegged to its initial value
$\Gamma_i$. Physically, this is so because, a such a (incorrect) dust solution is
necessarily radiationless,
and comparable to free
fall of a test particle, for which $\Gamma=\Gamma_0= conserved~ energy~ per~
unit~ rest~ mass$. And there is no question of $\Gamma$ assuming a negative
value in such a case:
In this case the final state could be marked with
\begin{equation}
 \Gamma_f^2=U_f^2 >0,   \qquad V_f^2=1, \qquad 2 M_f/R_f=1.
\end{equation}
In Paper I,
we have seen that Oppenheimer-Volkoff (OV) equation\cite{38} demands that for such dust
solutions $\rho\equiv 0$, and consequently, not only $M_f=0$, but also
$M_i=0$. This is possible in a strict sense only if $N\equiv 0$. Thus we
shall not consider this case any further.

\subsection{Static  Final States With $ 2M/R <1$}
Static final states will be marked by $V_f=U_f=0$
with  $\Gamma_f >0$:

\begin{equation}
\Gamma_f^2 >0, \qquad U_f=V_f=0, \qquad 2 M_f/R_f <1
\end{equation}

\subsection{Static Final State With $2M/R=1$ }

\begin{equation}
 \Gamma_f^2=U_f^2 =0, \qquad  V_f=0, \qquad 2 M_f/R_f =1
\end{equation}

\subsection{Dynamic Final State With $2M/R =1$}
A dynamic final state will have $V_f >0$. But since $V_f$ is allowed to be
non-zero in this case, the {\em final state} must be marked with the
maximum possible value of $V_f =1$:

\begin{equation}
 \Gamma_f^2=U_f^2 =0,   \qquad V_f^2=1, \qquad 2M_f/R_f=1.
\end{equation}

\section {More on the Physical Solutions}
We may note that
all the physical  solutions correspond to $U_f\rightarrow 0$.
A physically meaningful final state must have $U_f=0$ because of the
simple the fact that the final state, by definition, must correspond to an
extremum of $R$ and therefore, we must have $U_f=dR/d\tau = 0$.
However, it does not mean that the march towards the final state
would be monotonically
decelerated for a real fluid. Depending on the unpredictable actual solutions of the
complicated non-linear coupled collapse equations and the radiation
transfer mechanism, the system may even undergo phases of acceleration and
deceleration in the fashion of a damped oscillator. In contrast,  a
conventional (incorrect) dust-collapse picture  would
suggest monotonic acceleration and no emission of radiation. In such a picture,
a neutron star (NS) would be born on a free fall time scale
$<1$ ms. But we know that actual microphysics intervenes in such a case
and ensures that the $\nu$
signal heralding the birth of the NS is dictated by radiation diffusion
time scale, which, in this case is $\sim 10$s.

And
since it is $U$ and not $V$ which appears in the collapse equations, in a
sense, the final state here corresponds to the static Oppenheimer-Volkoff limit:
\begin{equation}
\partial R/\partial r \rightarrow 1; \qquad R\rightarrow 0,
\end{equation}
\begin{equation}
-g_{\rm rr}^{1/2} =e^\lambda ={1\over \Gamma} {\partial R\over \partial r}
\rightarrow {1\over \Gamma},
\end{equation}
\begin{equation}
{d M\over dR} \rightarrow 4\pi R^2 \rho; \qquad R\rightarrow 0,
\end{equation}
and
\begin{equation}
{dp\over dR} \rightarrow -{(\rho+p)(4\pi p R^3 +M)\over \Gamma R^2}.
\end{equation}
Note, however, that the $U=0$ case truly corresponds to a {\em locally static}
case only as long as $V_f= 0$, such as cases V.B and C.

In general, we see that
that $\Gamma_f \ge 0$, and $\Gamma_f$  cannot assume any
negative value. This conclusion could have been directly drawn from Eq.(5)
as well. If the fluid is collapsing (expanding), on physical grounds both $U$ and $V$
must be negative (positive), implying $\Gamma \ge 0$. Also by using the general condition that
$M_f \ge 0$, it can be shown that $\Gamma_f \ge 0$.

\subsection{Final State With $2M/R <1$}
The case IV.B  corresponding to $2GM_f/R_f <1$ with $V_f =0$ indicates the
formation of
static Ultra Compact Objects (UCOs) of  finite size $R_f$. An UCO may be
considered as a special case of ECO with $V_f=0$. We shall show
shortly that if one insists for a $R_f=0$  solution in this,
the configuration would have zero baryon number. Thus this solution indeed
corresponds to finite size static UCOs.

 To some extent, the  solutions  for this case  have already
been discussed in the literature\cite{38,39}. OV
used ideal Fermi-Dirac EOS and sought solutions with central density
$\rho_c=\infty$ and central pressure $p_c=\infty$:
\begin{equation}
\rho \rightarrow {3\over 56 \pi R^2},
\end{equation}
\begin{equation}
p\rightarrow {1\over3} \rho,
\end{equation}
\begin{equation}
M\rightarrow {3\over 14 G} R,
\end{equation}
\begin{equation}
\Gamma=constant=(4/7)^{1/2}
\end{equation}
Here it should be remembered that, as long as $R(t)\neq 0$, i.e., when one is
dealing with a non-singular configuration which may nevertheless harbour a
singularity at the center, the usual boundary condition that (i) $p=0$, at
$R=R_0$, the external surface, must be honoured. And only when a complete
singular state is reached with $R_0 =0$, the solution must be allowed to
be truly discontinous in $p$ and $\rho$. This shows that a strictly
correct solution of the above referred OV solution could be of two types:
(1) $R_0 =\infty, ~M(R_0)=\infty$, which is an unstable solution except
for photons; (2) $R_0=R=0, ~M(R_0)=0$, and this the only stable solution
for baryons. This conclusion also follows from the solution of Misner \&
Zapolsky\cite{39} which shows that $M_{\rm core} =0, ~R_{\rm core} =0$ when
$\rho_c$ is indeed $\infty$.
And, though,
the latter OV solution is a stable one, it is devoid of any physical
content because, as was specifically discussed by OV, it has $N=0$. This
can be verified
in the following way. Note that since luminosity $L\rightarrow 0$ in the
static limit, we
must have $p_{\rm radiation} \rightarrow 0$ as $R\rightarrow 0$, so that
the total pressure $p=p_{\rm matter}$ in this limit. In the low density
limit degenerate fermions have an EOS $p \propto n^{5/3}/ m$. Since this
EOS depends on the value of $m$, it is possible that in the low density
limit, it depends on the specific nature/variety of the fermions. In
contrast, as $\rho \rightarrow \infty$, the fermi EOS becomes independent
of $m$
\begin{equation}
p =0.25 \sqrt{3/8\pi} h c n^{4/3}
\end{equation}
where $h$ is Planck's constant. In fact, it may be shown that black body photons tend to obey the $p
\propto n^{4/3}$ EOS and it is likely that all extremely degenerate bosons with a finite value
of $m$ tends to follow this EOS in the limit $\rho/m \rightarrow \infty$.
In any case for the fermions, Eqs.(70), (71), and (74) would suggest that
\begin{equation}
n=n_0 R^{-3/2}
\end{equation}
where $n_0$ is an appropriate constant. Note that for static solutions,
 by using Eqs.(67) and (72), we have
\begin{equation}
N=  \int_0^{R_0} 4\pi n e^{\lambda} R^2 dr ={4\pi
n_0}\int_0^{R_0} {R^{1/2}\over \Gamma} dR
\end{equation}
 From this Eq. it is seen that for a constant $\Gamma\neq 0$,
$N\rightarrow 0$ as
$R_0\rightarrow 0$. It can be verified that,
even if we used a more general form of Eq.(74) like $p \propto
n^{4+\alpha\over 3}$, for the $\Gamma_f >0$ case all singular solutions
have $N=0$ if $\alpha <1$. Therefore as far as final states with $R_f=0$
are concerned, we discard the $2GM_f/R_f <1$ case
as physically meaningless. This conclusion corroborates the spirit
of the Cosmic Censorship Conjecture that spherical collapse cannot give
rise any naked singularity. As mentioned before many numerical/semi-analytical
studies on inhomogeneous dust collapse find the absence of formation of EH
in accordance with our exact result. And this absence of EH is interpreted
in terms of formation of ``naked singulaities''. In a strict sense
occurrence of true gravitational singularity would mean that the proper
radial distance measured along radial world lines $l =finite$. It is
likely that such numerical/semi-analytical studies implicitly assume that
indeed $l=finite$ and thus use the term ``singularity'' in this context.
But we shall see shortly that actually $l=\infty$ if the collapse would
proceed upto $R=0$ and therefore eventhough there is a ``nakedness''
(non-occurrence of EH), there would actually be no true singularity.

However we remind that if $R_f >0$,
$2M_f/R_f <1$ is a physically valid configuration with finite $N$ (UCO).

\subsection{Solutions With $2M_f/R_f =1$, $M_f=0$, $R_f=0$}
As far as true singular states are concerned,  the meaningful solutions
are V. C and D
 with $\Gamma_f =0$, because, for
a physical fluid, this null value of $\Gamma_f$ implies that $M_f=0$ even
though $M_i >0$. And out of these two probable cases, the solution D
 is of
particular astrophysical importance. Since in this case, $V_f\rightarrow constant (=1)$
for all fluid elements, it signifies complete kinetic ordering of the
fluid and, hence, eventual zero entropy. On the other hand, solution C
might correspond to a locally static $(V_f=0)$ configuration where all the
fluid particles form a single coherent perfectly degenerate system. In
this case too, the entropy of the system would be $\sim \ln(1) =0$.

With reference to the cases V.C and D,  by inspection, we have found that the
following class of solutions are suitable:
\begin{equation}
\rho \rightarrow {1\over 8\pi G R^2}; \qquad R\rightarrow 0, \qquad
U\rightarrow 0,
\end{equation}
\begin{equation}
p\rightarrow {a_1   \over 8\pi G R^2}; \qquad a_1>0,
\end{equation}
\begin{equation}
{2 G M\over R} \rightarrow 1 -\epsilon A R^{1+2 a_2};  \qquad a_2 <0.5,
\end{equation}
where $\epsilon \ll 1$ is a tunable constant. Here the normalization
constant
\begin{equation}
A= R_0^{2-2 a_2},
\end{equation}
and,
\begin{equation}
\Gamma \rightarrow \left(1-{2G M\over R}\right)^{1/2} \rightarrow \sqrt
{\epsilon A} R^{a_2 +0.5},
\end{equation}
so that,
\begin{equation}
M \rightarrow {R\over 2G}.
\end{equation}
By using Eqs.(76) and (81), we find that
\begin{equation}
N={4\pi n_0\over \sqrt{\epsilon A}} \int_0^{R_0} R^{-a_2} dR
={4\pi n_0\over \sqrt{\epsilon}(1-a_2)}
\end{equation}
Now by tuning (decreasing) the value of $\epsilon$ one can accommodate arbitrary number
of baryons in the singular state!
Recall that the for the external Vaidya metric, the dynamical collapse solution obeys the following boundary
condition at the outer boundary\cite{12,14}: $\sqrt{-g_{00}}=e^\psi= \Gamma +U$
in order that $\psi \rightarrow 0$ as $R\rightarrow \infty$, i.e, to have
an asymptotically flat spacetime. Then as $R_0\rightarrow R\rightarrow 0$
and $U\rightarrow 0$, we find that
\begin{equation}
e^\psi =\sqrt{-g_{00}} \rightarrow 0
\end{equation}
This shows that the {\em total surface red-shift $z\rightarrow \infty$
even though $M\rightarrow 0$}. Similarly, by using Eq.(79),  we find that,
the typical curvature component
\begin{equation}
R^{\vartheta \varphi}_{\vartheta \varphi} ={2G M\over R^3} ={1\over
R^2}\rightarrow \infty
\end{equation}
as $R\rightarrow 0$.
Also, using Eq.(81), note that the proper volume of the fluid:
\begin{equation}
\Omega =\int_0^{R_0} {4 \pi R^2 \over \Gamma} dR \sim {1\over\sqrt{\epsilon}}
R_0^{1.5}\rightarrow 0
\end{equation}
since $\epsilon$ is finite.
Nevertheless, most importantly, by using Eqs.(66), (67), and (81), it
follows that, the proper radial distance to be travelled to reach this singular
state (along a radial worldline) is
\begin{equation}
Lt_{R_0\rightarrow 0} l \ge \int_0^{R_0} {dR\over \Gamma}\sim {1\over\sqrt{\epsilon}}
R_0^{-0.5}
\end{equation}
Even if $\epsilon \neq 0$, i.e., even if $N$ is finite, we see that $l \rightarrow \infty$ as $
R_0 \rightarrow 0$!  Note that when the metric components are dependent on
time, in general, $l$ cannot be meaningfully defined\cite{15}, but here we are
defining $l$ only for a specific limit.  Correspondingly, the proper time (as measured
following the same radial worldline and defined in the same specific limit) required to attain
this state would also be infinite
\begin{equation}
\tau = Lt_{R_0\rightarrow 0} \int {dl\over v} \ge {l\over c} \rightarrow \infty
\end{equation}
This means that time like  radial worldlines are never terminated
and
{\em the fluid can never
manage to attain this state of singularity} although it can strive to do
so continuously.  Now recall that, the most ideal condition for formation
of a BH or any singularity is the spherically symmetric collapse of a
perfect fluid not endowed with any charge, magnetic field or rotation.
If no singularity can ever be formed  under this most
ideal condition, it is almost certain that any other realistic and more
complicated case of collapse of physical matter will be free of singularities. This means
that as far as dynamics of isolated bodies are concerned  GTR may indeed be
free of any kind of singularities (this does not rule out Big-Bang type singularities).
 Moreover, if a BH were formed its zero
surface area  would imply zero entropy content.
Thus spherical gravitational collapse does not entail any loss of
information, any violation of baryon/lepton number conservation (at the
classical non-quantum level, i.e, excluding CP violation etc. etc.) or any other
known laws of physics. The {\em fluid simply tries to radiate away its
entire mass energy and entropy in order to seek the ultimate classical
ground state} $M=0$.

\section {Summary and Discussions}
So far there have been hundreds of claims for having found observational
evidences for BHs. But from several independent approaches and from most
basic premises, our work has shown that GTR does not allow existence of
finite mass BHs. The immediate question would be then what is the true
nature of the observationally discovered BH candidates (BHC). Note that
often these claims mean hint for detection of compact objects with masses
larger than the upper limit of the mass of {\em stable and cold} Neutron
Stars (NSs) and our work too suggests existence of massive compact
objects. However our work also suggests that such massive objects need not
be {\em cold and static} and neither can they have any EH.

   There are, nonetheless, claims  for the ``detection'' of
EHs in some BHCs\cite{41,42}. The argument in favor of the claim
is broadly the following: At very low accretion rates, the coupling between
electrons and ions could be very weak. In such a case most of the energy
of the flow lies with the ions, but since radiative efficiency of ions is
very poor, a spherical flow radiates insignificant fraction of accretion energy
and carries most of the energy towards the central compact object. Such a
flow is called Advection Dominated Flow (ADAF). If the central object has
a ``hard surface'', the inflow energy is eventually radiated from the hard
surface. On the other hand, if the central object is a BH, the flow energy
simply disappears inside the EH. For several supermassive BHCs and stellar
mass BHCs, this is claimed to be the case.
But   there could be several caveats in this interpretation:

 (i) Garcia et al.\cite{41} based their entire claim of having ``detected'' EH on a
number of assumtions one of which is that X-ray binaries with similar period should have
similar quiesent mass accretion rates. But we know that several compact
X-ray binaries like Cygnus X-3 with orbital period $P=4.8$ hr and Sco X-1
(P=19 hr) accrete mass at the Eddington rate. So, if we follow the
argument of Garcia et al. a large number of BHC X-ray binaries  with P
ranging from 5.1 (GRO J0422+32) -27.0 hr (4U 1543-47) should have had near
Eddington accretion rate too.  But the estimated accretion rates in such
cases  are $10^{-6} - 10^{-7}$ Eddington rate. Thus clearly the
fundamental assumtion of Garcia et al. is incorrect. The mass accretion
rate, depends not only on the value of P but more importantly on factors
such as the nature and evolutionary state of the mass donating companion.
And in any case,  for such compact binaries,  if there is a
viable mass tranfer mechanism, the accretion rates are likely to be much higher.

(ii)
Such low luminosities may not be due to accretion at all. On the other
hand such low luminosities could be
 due to Magnetospheric emission of spinning Neutron Stars or BHCs.
Recently, Robertson\cite{3} and Robertson and Leiter\cite{4}
 have  {\em explained the X-ray
emission from several BHCs having even much higher luminosities as
magnetospheric origin associated with their strong intrinsic magnetic
field}. Note that, if the BHCs were really BHs, they would not have had
any intrinsic magnetic field whereas if they are Eternally Collapsing
Objects (ECOs) or static Ultra Compact Objects (UCOs) with physical
surface they are expected to have such strong intrinsic magnetic fields.
Campana and
Stella (2000)\cite{43} have also argued that part of the observed emission from the BHCs
could be of magnetospheric origin rather than due to accretion.
 Vadawale, Rao and Chakrabarti\cite{44} have explained one
additional component of hard X-rays from the micro-quasar GRS1915+105 as
Synchrotron radiation.

(iii) The accretion luminosity may be radiated not only in soft X-rays but,
in certain cases, also in the form of $\gamma$-rays or neutrinos. In other
words, the luminosity in the soft X-ray band need not be proportional to
the total mass accretion rate (Campana and Stella 2000)\cite{43}.

(iv) The x-rays if assumed to be of accretion origin, could be coming from
an accretion disk or a corona and not from a spherical flow.

The
major part of the accretion flow might also be rebounded in the form of ``jets''
rather than in the form of any radiation not envisazed in the original
ADAF model

  In a more recent review paper Narayan, Garcia and McClintock (2001)\cite{42}
  have admitted that some of the assumptions behind the ADAF model may be
incorrect (although on the basis of which repeated claims have been made
about the ``detection'' of EHs in BHCs). They  admit that various MHD
processes (which are hardly understood) may affect the flow dynamics and
the flow may be Convection dominated rather than Advection dominated.
 This only shows that the claims of ``detection'' of EH are based on
poorly understood physics.

(v) Some of the BHCs in the X-ray binaries may be spinning with a period
of few milliseconds. They are likely to be aspherical and potential
sources of gravitational radiation. Accretion can cause such objects
wobble more and radiate more gravitational radiation.

(vi) In extremely strong gravitational field associated with some BHCs ($z \gg 0$), overall
accretion physics may be somewhat different from what we expect from
studies of NSs ($z \sim 0.1$). In certain cases both
 accreted matter and energy might
get ``stuck'' by extremely high gravity atleast for certain period of time
and which might be reemitted later. Menou has also urged that claims of detection
of ``BH'' cannot be made just on the basis of difference in luminosities in two
classes of sources\cite{45}. Recently Narayan and Heyl\cite{46} have
offered yet another evidence for detection of ``EH': they mention that
while some of the NS X-ray binaries show Type I X-ray burst (nuclear flash
on the physical boundary of the compact object) none of the BHCs show such
bursts. Here they have ignored the fact that the presence of intrinsic
magnetic field considerably higher than $\sim 10^{9}$ G may suppress Type
I burst activities in the UCOs. Thus such an absence of Type I bursts
actually enhances the case that stellar mass BHCs may possess strong
intrinsic magnetic field\cite{3,4}.

Therefore there is really no rigorous evidence for ``detection'' of any EH contrary
to the repeated claims made to this effect. On the other hand, the
evidence for presence of intrinsic magnetic field in the BHCs\cite{3,4}
suggests that the BHCs are not BHs.
 As we show, continued collapse of sufficiently massive
objects may continue indefinitely because
{\em the ever increasing curvature of spacetime (components of Ricci Tensor) tends to stretch
the physical spacetime to infinite extent}.
  Therefore, it is
probable that, subject to the presently unknown behaviour of the arbitrary
high density nuclear EOS and unknown plausible phase transitions of
collapsing matter under such conditions, GTR gravitational collapse may find
one or more quasi-stable ultracompact static or even dynamically
collapsing configurations.
Such objects, more massive and more compact than canonical Neutron
Stars (NSs)  have been broadly called
Eternally Collapsing Objects (ECO).
 If in a given epoch the gravitational
mass of a BHC/ECO is $M$, its circumfernce radius $R$ can be arbitrarily close
to its instantaneous Sch. radius $R_g= 2GM/c^2$ without ever becoming less than $R_g$,
i.e, $R(t) \ge R_g(t)$. For practical purposes such objects are as compact as a
supposed BH and they would satisfy many of the ``operational definitions''
of BHs.
But whereas the EH is characterized by a gravitational redshift $z=\infty$
such BHCs or ECOs have finite $z$. In case the ECO is very massive and
dynamically collapsing it might try to attain the $z=\infty$ state after
infinite proper time (as measured along radial worldlines) by which time
its mass would be $M=0$.
Thus these objects  need not always be
static and cold and they need not represent stable solutions of equations for
hydrostatic balance.  They may be
collapsing  with substantial local speed $V$, (which this work cannot predict)
 but the speed of collapse
perceived by a distant observer ($V_\infty$) would approximately be lower by a factor
of $(1+z)^2$.  Since $V$ is finite ($<c$) and eventually $z \rightarrow
\infty$, the ultimate value of $V_\infty \rightarrow 0$. So it is likely that
even for accurate measurements (which might be possible in remote future)
spanning few years, an {\it isolated} ECO may appear as a ``static'' object.
The value of $R$ for an accreting ECO would decrease even more slowly.
 Thus gravitational collapse of sufficiently massive bodies should indeed
result in objects which could be more compact and arbitrarily more massive
than typical NSs ($z > \sim 0.1)$.
It is found that, if there are anisotropies, in principle there could be
static objects with arbitrary high (but finite) $z$\cite{47}. Even within the
assumption of spherical symmetry, {\em non-standard QCD may allow existence of
cold compact objects with masses as large as} $10 M_\odot$ or higher\cite{48}.

Such stars are
called Q-stars (not the usual quark stars), and they could be much more compact than
a canonical NS; for instance,
a stable  non rotating Q-star of mass 12$ M_\odot$ might have a radius of
$\sim 52$ Km. This may be compared with the value of $R_{gb} \approx
36$ Km of a supposed BH of same mass.
 And, in any
case, when we do away with the assumption of ``cold'' objects and more importantly,
staticity condition there could be objects with arbitrary high $z$.

Recently, Mazur and Mottola\cite{2} have suggested that when the EH would
be about to be formed during collapse, the quantum back reaction due to
extremely strong gravitational field would first cause a phase transition
in the collapsing matter such that (i) $p \rightarrow \rho$ and then
(ii) $p\rightarrow -\rho$. If this would happen, the collapse would be halted
and there could be static UCOs with arbitrary high mass and $z$. Following
the theory of relativistic polytropes, we have found that even if one
avoids the unusual phase transition (ii) and restricts oneselves to (i) or
any EOS with $p = \beta \rho$, with $\beta \sim 1$, it is possible to have
UCOs of arbitrary mass but a $z <2$. In particular,  for
stellar mass objects, if $\beta=1$
 and such a phase transition takes place at nuclear
density $\sim 2. 10^{14}$ g/cm$^3$, the maximum mass of the UCOs would be
$\sim 11 M_\odot$ which happens to be upper limit for the mass  of the
stellar mass BHCs\cite{49}.

 Recall here that there are many globular clusters which are believed to
have undergone core collapse and certainly harbour massive BHs. But all
persistent  observations have failed to detect the presence of any such
BHs\cite{50,51}.  And as far as the active galactic nuclei are concerned, it
is already a well known idea that their centers may contain supermassive
stars at various stages of contraction\cite{33,52,53} (which can very well accrete
surrounding matter) or dense regions of star
bursts. However, Newtonian or Post Newtonian models of
supermassive stars cannot explain the sustenance of galactic nuclei for
periords longer than few years. On the other hand, we have shown here
that, actual Relativistic Configurations (Supermassive stars or stellar
mass ECOs) may appear to be almost static for distant observers for any
amount of finite duration and yet {\em keep on contracting internally} with
finite value of $V$. There are some tentative evidences that the supposed
supermassive BHs at the core of many galaxies may not be BHs (as confirmed
by the present study):

 The center of our galaxy  harbours a BHC, Sgr A$^*$, of mass
$2.6 \times 10^6 M_\odot$. The recent observation of $\sim 10-20\%$ linear polarization
from this source has strongly suggested against ADAF model\cite{54}.
 On the other
hand, the observed radiation is much more likely due to Synchrotron
process\cite{54}.
In fact, even more recently, Donato, Ghisellini \& Tagliaferri\cite{55}
 have shown that
the low power X-ray emission from the AGNs are due to Synchrotron process rather
by accretion process.

 Munyaneza \& Viollier\cite{56} have claimed that the accurate studies of the
motion of stars near Sgr A$^*$ are more amenable to a scenario where it is
not a BH but a self-gravitating ball of Weakly Interacting Fermions of
mass $m_f >\sim 15.9$ keV. Recall that the OV mass
limit may be expressed as
\b
M_{OV} = 0.54195 M_{pl}^3 m_f^{-2} g_f^{-1/2} = 2.7821 \times 10^9 M_\odot
(15 keV/ m_f)^2 (2/g_f)^2
\e
where $M_{Pl}= (\hbar c/G)^{1/2} $ is the Planck mass and $g_f$ is the
degeneracy factor. With a range of  $ 13<m_f <17 $ keV, these authors
point out that the entire range of supermassive BHCs can be understood.
Bilic\cite{57} has also suggested that the BHCs at the centre of galaxies
could be heavy neutrino stars. Svidzinsky\cite{58} has suggested that atleast some
of the BHCs in the blazars could be heavy bosonic stars.

Note that the progenitors of the ECOs or BHCs must be much more massive (and larger in size)
than those of
the NSs.
 Then it
 follows from the magnetic flux conservation law that BHCs (at the galactic level)
 should have
magnetic fields considerably higher than NSs. It is also probable even
when they are old, their diminished magnetic fields are considerably
higher than $10^{10}$G. In such cases, BHCs will not exhibit Type I X-ray
burst activity. There may {\em indeed be evidence for
 intrinsic (high) magnetic fields} for the BHCs\cite{3,4}.
 However, in
some cases, they may well have sufficiently low magnetic field and show
Type I bursts. It is now known that Cir X-1 which was considered a BHC,
did show Type I burst, a signature of ``hard surface''. Irrespective of
interprtation of presently available observations, our work has shown that the
BHCs cannot be, in a strict sense, (finite mass) BHs because then {\it timelike
geodesics would become null} on their EHs.

Although future observations must settle questions
about the true nature
of BHCs,
we remind the reader, that, {\em our results are exact subject to the validity of
GTR}. Finally, if we demand that at sufficiently small  scales quantum mechanics
must take over GTR, the final state of continued ideal spherical collapse
might be a Planck glouble of mass $M_{\rm pl} \sim 10^{-5}$g,
irrespective of the initial value of $M_i$. We can only conjecture now
whether the globule may or may not comprise wiggling elementary strings.

   There have been many publications which claim that studies of
gravitational collapse leads to formation of either BHs or naked singularities.
And some readers may wonder why our result differs from so many studies.
Here, first, it should be borne in mind that all such studies involve
either semi-analytical or semi-numerical treatment. And since pursuance of
singularity, by definition, involve handling of infinite values of various
physical quantities, only, the results obtained by exact analytical
treatments could be reliable. Again as asserted in the introduction, one
must be able to handle the progress of the exact ever evolving EOS and
radiation transport properties all the way upto the singularity in any
analytical or numerical study in case that study requires specific form of
EOS or radiation transport properties. Now obviously we cannot treat
situations involving arbitrary large  pressure, radiation density etc in
an exact manner both because of theoretical and numerical limitations.
Thus, in a strict and true sense, study of the problem of
gravitational collapse of physical matter by any brute force direct method of ``solving'' is
out of question. And if we take stock of the events, perhaps, (apparently)
was the attempt by OS. And as we saw the result that trapped surfaces do
not form is also hidden in Eq.(36) of their paper. And as mentioned earlier,
the fluid considered by OS did not correspond to any physical fluid
because all physical fluids have finite pressure and density gradient
(under self-gravity). Finally the {\em must be allowed to radiate out} as it collapses
otherwise its gravitational mass $M$ would remain fixed and a trapped
surface might appear.

Even then, we feel that if the authors of all such papers would (i) either
    try to find a global equations (5) and (7), or (ii) write an exact
equation (such as Eq.36 of OS) connecting $T$ with the Invariant
circumference Radius (a scalar) $R$, they would arrive at the
conclusion same as ours. In other words, before attempting to ``solve''
the collapse equations, if they would study their generic properties, they
would also find that the final state would correspond to $M=0$. And as
pointed in Paper I, even though  the previous authors overlooked such
crucial generic properties, atleast in some cases there are broad or even
exact agreements with our results. For instance

(i) Our result that there are no finite mass BHs is (apparently)
in agreement with the finding that singularities could be ``naked'' and
not finite mass BHs.

(ii) Our result that gravitational singularities should have $M=0$ is in
agreement of the result of Lake\cite{59}. As mentioned in Paper I, this is
also in agreement with several other previous ideas\cite{60,61}.

(iii) Our assertion that realistic studies of gravitational collapse of
physical matter must incluse pressure gradient is shared by many authors.
Specifically,  we would like to mention the work of Cooperstock, Jhingan,
Joshi and Singh (1997)\cite{62}. In this work, these authors pointed out that
inclusion of pressure gradient forces is very important and formation of
BH may be avoided by it. They also found that ``the naked singularity
arising in spherical collapse is necessarily massless in the sense that
the mass function $m(t,r)$ will go to zero at the naked singularity''  in
conformity with our Eq.(16) (their $m$ is ours $M$). Below Eq.(17)
of this paper, these authors
write that

``in the limit $R\rightarrow 0$ we have $m=R/2$''

 in exact
conformity with our result that the final state corresponds to a zero mass
BH. But since the very concept of a BH generally implies $M >0$, these
authors called such a final state to be ``marginally naked'' rather than a
Black Hole.  In the context of studies
of final state of spherical gravitational collapse, these authors correctly admit that,
 that ``most of the results available thus far numerical in nature''.

 One of the definitive signature of existence of finite mass BHs would be the
 evaporation of primordial BHs at the present epoch\cite{63}. Gamma ray astronomers
have been searching for these phenomenon for the last 25 years without
any success.
 And we can predict
with absolute certainty that no such events would ever be detected because
GR does not allow formation/existence of finite mass BHs.

 If a BH would exist, the proper
length of an infalling astronomer or anything would become $\infty$ as it
would approach the central singularity\cite{23} because of infinite tidal gravity.
 Then how would the observer
 stay
put in a geometrical point? There is no resolution
 for such inconsistencies within the BH paradigm.
 Such inconsistency actually does not arise because we
have shown that (1) there cannot be any finite mass GR singularity for
isolated bodies and (2) in case the astronaut
 is falling along with the collapsing matter, the proper radial length
 of not only the astronaut but that of the star too
 tend to become infinite. And there would be
no problem in accommodating an infinitely stretched astronaut in the
collapsing body.

Note added in proof: After this paper was sent to press, Dr. Torres of Princeton Univ.
pointed out that it is likely  that the BHC at the galactic center and several other BHCs
could be Bosonic stars. For all practical purposes the compactness of these bosonic stars
could be like that of supposed BHs. Further, since both neutrino condensations or such bosonic
condensations comprise weakly interacting particles, matter accreting onto them would hardly
interact on impingement. Thus even though such condensations will have a physical surface,
it may appear that matter is being lost through an ``Event Horizon''.

Ref:

1. D.F. Torres, S. Capozziello, and G. Lambiase, {\it Phys. Rev.}, {\bf D62}, 104012, (2000),
astro-ph/0004064.

2. D. Torres, {\it Nucl. Phys.}, {\bf B 626}, 377, (2002), hep-ph/0201154.
\end{document}